\documentclass{jfm}

\usepackage{amsmath}
\usepackage{color}
\usepackage{amssymb}
\usepackage{graphicx}
\usepackage{upgreek}
\usepackage{textcomp}
\usepackage{siunitx}

\newcommand{\eqr}[1]{(\ref{eq:#1})}

\newcommand{\ods}[2]{\frac{d^2 #1}{d #2^2}}

\newcommand{\ie}{\textit{i.e.}}
\newcommand{\bs}[1]{\boldsymbol{#1}}

\graphicspath{ {./} }

\usepackage{dcolumn}% Align table columns on decimal point
\usepackage{bm}% bold math
\usepackage{amsmath}
\usepackage{amssymb}
\usepackage{scalerel}[2014/03/10]
\usepackage{stackengine}

\usepackage[euler]{textgreek}

\usepackage{multirow}

\title[ ]{A model to predict the oscillation frequency for drops pinned on a vertical planar surface}

\author{J. Sakakeeny\aff{1},
C. Deshpande\aff{2}, 
S. Deb\aff{2},
J. L.~Alvarado\aff{2,3}, 
Y. Ling\aff{1} \corresp{\email{stanley\_ling@baylor.edu}}
}

\affiliation{
\aff{1}Department of Mechanical Engineering, Baylor University, Waco, TX 76643, USA
\aff{2}Department of Mechanical Engineering, Texas A\&M University, College Station, TX 77843, USA
\aff{3}Department of Engineering Technology and Industrial Distribution, Texas A\&M University, College Station, TX 77843, USA
}

\begin{document}

\maketitle

\begin{abstract}
Accurate prediction of the natural frequency for the lateral oscillation of a liquid drop pinned on a vertical planar surface is important to many drop applications. The natural oscillation frequency, normalized by the capillary frequency, is {mainly} a function of the equilibrium contact angle and the Bond number (Bo), when the contact lines remain pinned. Parametric numerical and experimental studies have been performed to establish a comprehensive understanding of oscillation dynamics. {An} inviscid model has been developed to predict the oscillation frequency for wide ranges of Bo and contact angles. The model reveals the scaling relation between the normalized frequency and Bo, which is validated by the numerical simulation results. For a given equilibrium contact angle, the lateral oscillation frequency decreases with Bo, implying that resonance frequencies will be magnified if the drop oscillations occur in a reduced gravity environment. 
\end{abstract}

%Use showkeys class option if keyword
                              %display desired

\section{Introduction}
Effective drop removal from a condenser surface is critical to highly-efficient dropwise condensation, which is in turn important to many applications such as condensation heat transfer \citep{Yao_2017a} and water harvesting \citep{Dai_2018a}. The recent advancement on superhydrophobic surfaces \citep{Boreyko_2009b, Yao_2017a} have shown promising enhancement of droplet mobility. Nevertheless, high mobility of drops on superhydrophobic surfaces can be lost when the condensate nucleates within the texture and the wetting falls into the Wenzel state. To restore the high drop mobility, external forcing such as surface vibration is typically required to excite the oscillation of the drops and to revert the wetting to the Cassie state \citep{Boreyko_2009a, Yao_2017a}. When the excitation frequency matches the natural frequency, the oscillation amplitude of the drop is maximized for a given energy input  due to the resonance effect \citep{Noblin_2004a,Boreyko_2009a,Yao_2017a}. Therefore, it is highly advantageous to  predict the natural oscillation frequencies for drops on surfaces of different wettability.

\begin{figure}
 \centering
 \includegraphics[width=0.95\textwidth]{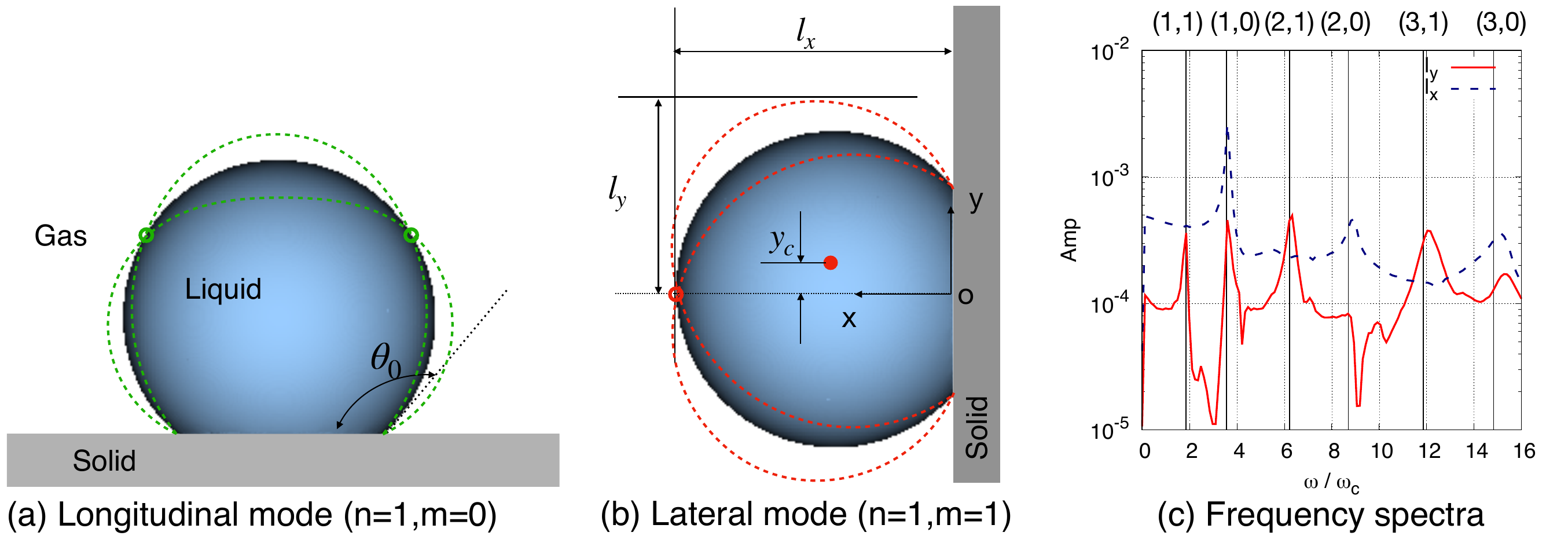}
 \caption{Schematics of the (a) longitudinal ($n=1$, $m=0$)  and (b) lateral ($n=1$, $m=1$) oscillation modes for a drop on a flat surface with pinned contact lines. The red and green open circles indicate the longitudinal and azimuthal nodal lines, while the red filled circle indicates the centroid of the deformed drop. (c) Simulation results for the frequency spectra for the maximum interfacial locations $l_x$ and $l_y$, for $\theta_0=\ang{100}$ and $\text{Bo}=0$. The B-S model predictions \citep{Bostwick_2014a} for modes $(n,m)$ are shown (see vertical lines) for comparison.  }
 \label{fig:schem_mode}
\end{figure}

%General description of oscillation dynamics and key parameters
Similar to a free drop, the oscillation of a drop supported by a planar surface can be characterized by spherical harmonic modes, with $n$ and $m$ representing the longitudinal and azimuthal wavenumbers, respectively \citep{Strani_1984a,Bostwick_2014a}. Different from free drops, the two $n=1$ modes, corresponding to the centroid translation in the directions normal $(n=1,m=0)$ and tangential $(n=1,m=1)$ to the surface, trigger shape oscillations for supported drops. Since the natural frequency and damping rate increase with $n$ \citep{Lamb_1932a}, the $n=1$ modes typically dominate the oscillations of supported drops \citep{Strani_1984a, Noblin_2004a, Sakakeeny_2020a,Sakakeeny_2021a}. While the longitudinal mode $(n=1,m=0)$ is axisymmetric, see {figure~\ref{fig:schem_mode}(a)}, and has been extensively studied \citep{Strani_1984a, Strani_1988a, Noblin_2004a, Sakakeeny_2020a, Sakakeeny_2021a}, less attention has been paid to the lateral mode $(n=1,m=1)$, see {figure~\ref{fig:schem_mode}(b)}, {which is the focus of the present study.}

{In the limit of zero-gravity, different inviscid theoretical models \citep{Strani_1984a, Celestini_2006a, Bostwick_2014a} have been developed to predict the frequency of different spherical-harmonic modes, including the lateral mode of interest here. As the effect of gravity is characterized by the Bond number (Bo), the aforementioned inviscid models are strictly valid only in the limit of zero Bo, though the model predictions were observed to agree reasonably well with experiment for drops with small Bo \citep{Chang_2015a}. A simple capillary-gravity wave model has been suggested by \citet{Noblin_2004a} to include the effect of gravity on the oscillation frequency, but significant discrepancies with experiment was found for surfaces with large contact angles \citep{Yao_2017a}. Numerical simulations have been recently performed by \cite{Sakakeeny_2020a, Sakakeeny_2021a} to investigate the effect of Bond number on the axisymmetric zonal oscillation modes for supported drops. Their results indicated that the drop oscillation frequency varies significantly with Bo, even when the drop radius is smaller than the capillary length ($\text{Bo}<1$). Furthermore, they also found that the orientation of  the supporting surface, with respect to the gravity acceleration, also matters: while the drop oscillation frequency increases with Bo for sessile drops, the frequency decreases with Bo for pendant drops. To the knowlege of the authors, there is no former experimental or numerical studies in the literature that systematically address the effects of Bo and equilibrium contact angle on the oscillation of drops pinned on a vertical planar surface. In the present study, a combined numerical, experimental and theoretical analysis is performed for the lateral oscillation of a water drop on a vertical planar surface. The goal of the study is to establish a model to predict the oscillation frequency for drop at small but finite Bo. 
We focus only on the drops with pinned contact lines, since they are the ones that require oscillation excitation to restore high mobility. }

\section{Methods}

\subsection{Key parameters}
The shape oscillation of a liquid drop is mainly controlled by surface tension. The contact with the solid surface has a significant impact on the oscillation dynamics. The natural frequency of the drop oscillation depends on the equilibrium contact angle ($\theta_0$), contact line mobility, and Bo \citep{Bostwick_2014a, Sakakeeny_2020a, Sakakeeny_2021a}. The key dimensionless parameters and the ranges of values considered are summarized in Table \ref{tab:physParamNonDim_Paper3}. Due to the small gas-to-liquid density and viscosity ratios, $r=\rho_g/\rho_l$ and $m=\mu_g/\mu_l$, the effect of the surrounding gas is generally small. The effects of gravity and liquid viscosity, compared to surface tension, are characterized by the Bond and Ohnesorge numbers, $\text{Bo}=\rho_lgR_d^2/\sigma$ and $\text{Oh}=\mu_l/\sqrt{\rho_l \sigma R_d}$, where  $g$ and $\sigma$ are gravity, and surface tension, while $R_d=(3V_d/4\pi)^{1/3}$ is the radius based on the drop volume $V_d$. For millimeter-size water drops, Oh and Bo are {smaller than one}. As a result, the oscillation frequency $\omega$ generally scales with the capillary frequency \citep{Lamb_1932a}, $\omega_c= \sqrt{\sigma / (\rho_l R_0^3)}$, where $R_0$ is the radius of the equilibrium spherical-cap shape when gravity is absent,  and $R_0= R_d \left((2 + cos \theta_0) (1-cos \theta_0)^2/4\right)^{-1/3}$. 
The reason for using $R_0$, instead of $R_d$ is because $R_0$ better represents the radius of curvature. 

{While $r$, $m$, and Oh are fixed in the simulation}, the equilibrium contact angle $\theta_0$ is varied from $\ang{50}$ to $\ang{150}$ to cover a wide range of wettability, varying from hydrophilic to superhydrophobic surfaces.  For a given liquid, Bo can be varied by changing $V_d$ or $g$. In the present study, we keep $V_d$ fixed and vary $g$, yielding $0\le \text{Bo} \le0.78$. Except for very small Bo, the simulation results also well represent cases for drops with different $V_d$ under full gravity. {Though $\text{Bo}$ is smaller than unity for all cases considered, namely $R_0$ is smaller than the capillary length $\sqrt{\sigma/(\rho_l g)}$, it will be shown later that the effect of gravity on the oscillation frequency can be significant.
}

\begin{table}
  \begin{center}
  \begin{tabular}{ccccc} 
        r & $m$ & $\text{Oh}$ & $\theta_0$ & $\text{Bo}$\\
        $\rho_g / \rho_l$ & $\mu_g / \mu_l$ & $\mu_l / \sqrt{\rho_l \sigma R_d}$ & (\textdegree) & $\rho_l g R_d^2 / \sigma$  \\
         $0.0012$ & $0.01$ & 0.00239 & ${50}$ - $150$ & 0 - 0.78\\
  \end{tabular}
    \caption{Key parameters for simulations.}
\label{tab:physParamNonDim_Paper3}
    \end{center}
\end{table}

\subsection{Simulation methods}
The numerical simulations were conducted using the \emph{Basilisk} solver, in which the Navier-Stokes equations are solved by a finite-volume approach. The sharp interface is captured by the volume-of-fluid (VOF) method \citep{Scardovelli_1999a}. The balanced-force method is used to discretize the surface tension \citep{Popinet_2009a}. The height-function method is used to calculate the interface curvature \citep{Popinet_2009a} and to specify the contact angle at the surface \citep{Afkhami_2009a}. Details of the numerical methods and validation studies for the \emph{Basilisk} solver can be found in previous studies \citep{Sakakeeny_2020a, Sakakeeny_2021a} and the code website. 

\begin{figure}
 \centering
 \includegraphics[width=0.95\textwidth]{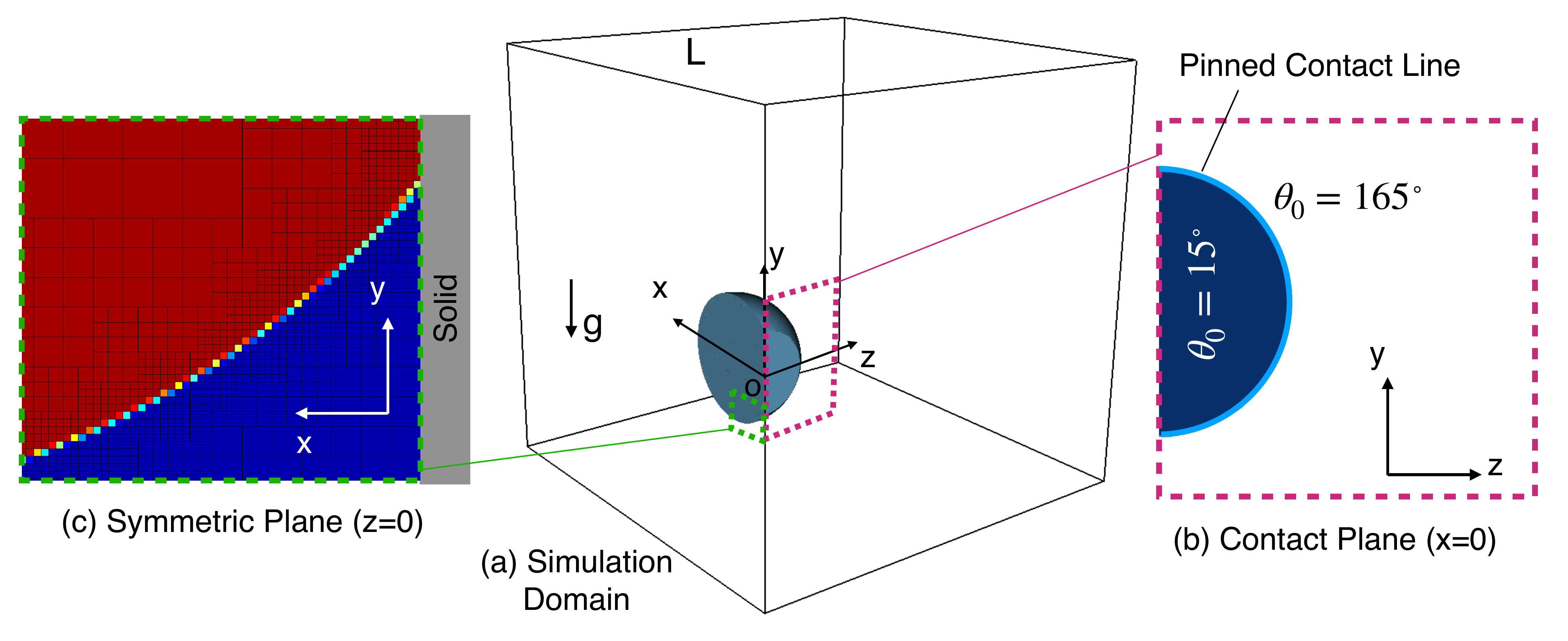}
 \caption{(a) The simulation domain. (b) {Closeup of the field of liquid volume fraction $f$ on the contact plane} ($x=0$), showing the pinned  contact line circle and the small and large $\theta_0$ specified inside and outside of the contact line. (c) Closeup of the symmetric plane ($z=0$), showing the adaptive mesh and VOF field. }
 \label{fig:sim_setup}
\end{figure}

The computational domain is a cube with the edge length $L=8R_d$, see figure~\ref{fig:sim_setup}(a). The drop is symmetric with respect to the plane $z=0$ for the lateral mode, so only half of the drop is simulated. The surface $x=0$ is in contact with the drop and is taken as a no-slip wall. Other boundaries are slip walls. Though modeling of mobile contact lines is challenging \citep{Afkhami_2018a,Snoeijer_2013a}, here we consider only the asymptotic limit that the contact lines are pinned. Physically this limit occurs when the hysteresis effect is very strong and the oscillation amplitude is small. The pinning of contact lines is achieved by imposing very small and large contact angles, \ie, $\theta_0=\ang{15}$ and $\ang{165}$, inside and outside of the contact line on the surface $x=0$, see figure~\ref{fig:sim_setup}(b), to mimic the strong hysteresis effect. As long as the contact angle $\theta$ remains bounded these two values when the drop oscillates, the contact line cannot recede or advance. Validation for of this treatment of contact lines can be found in our previous study \citep{Sakakeeny_2021a}. The drop oscillations are initiated by specifying an initial shape different from the equilibrium shape or by increasing the gravity for a short period. The oscillation frequency will not be affected if the amplitudes of excited oscillations are small. 

An adaptive octree mesh is used for spatial discretization, see figure~\ref{fig:sim_setup}(c), and the maximum refinement level is $\mathcal{L}=10$, corresponding to $R_d/\Delta_{\min}=128$, where $\Delta_{\min}$ is the minimum cell size. The adaptation thresholds for the liquid volume fraction {($f$)} and the fluid velocity components {($u,v,w$)} are 0.0001 and 0.01, respectively. It is verified that the refinement level and error thresholds are sufficient to yield mesh-independent results. The simulations were performed on the cluster \emph{Kodiak} using 144 CPU cores (Intel E5-2695 V4). Each simulation case takes about 15 days to reach the time $t\omega_c\approx 110$. The simulation time has been verified to be sufficiently long for accurate measurement of frequencies from the spectra. To vary both $\theta_0$ and Bo, there are in total about 30 cases simulated.

Figure \ref{fig:schem_mode}(c) shows the frequency spectra of the maximum interfacial locations in $x$ and $y$ directions, $l_x$ and $l_y$, for $\theta_0=\ang{100}$ and $\text{Bo}=0$. The different oscillation modes are clearly identified: while the spectrum for $l_y$ shows the $m=1$ modes, that for $l_x$ indicates the $m=0$ modes. The frequencies predicted by the inviscid model of \citet{Bostwick_2014a} (B-S) for $(n,m$) modes  are shown for comparison, see the vertical lines. The simulation results generally agree very well with the inviscid model, though small discrepancies were observed for the high-order (3,0) and (3,1) modes. It can be seen that the lateral mode considered in the present study exhibits the lowest frequency.

\subsection{Experimental methods}
\begin{figure}
 \centering
 \includegraphics[width=0.95\textwidth]{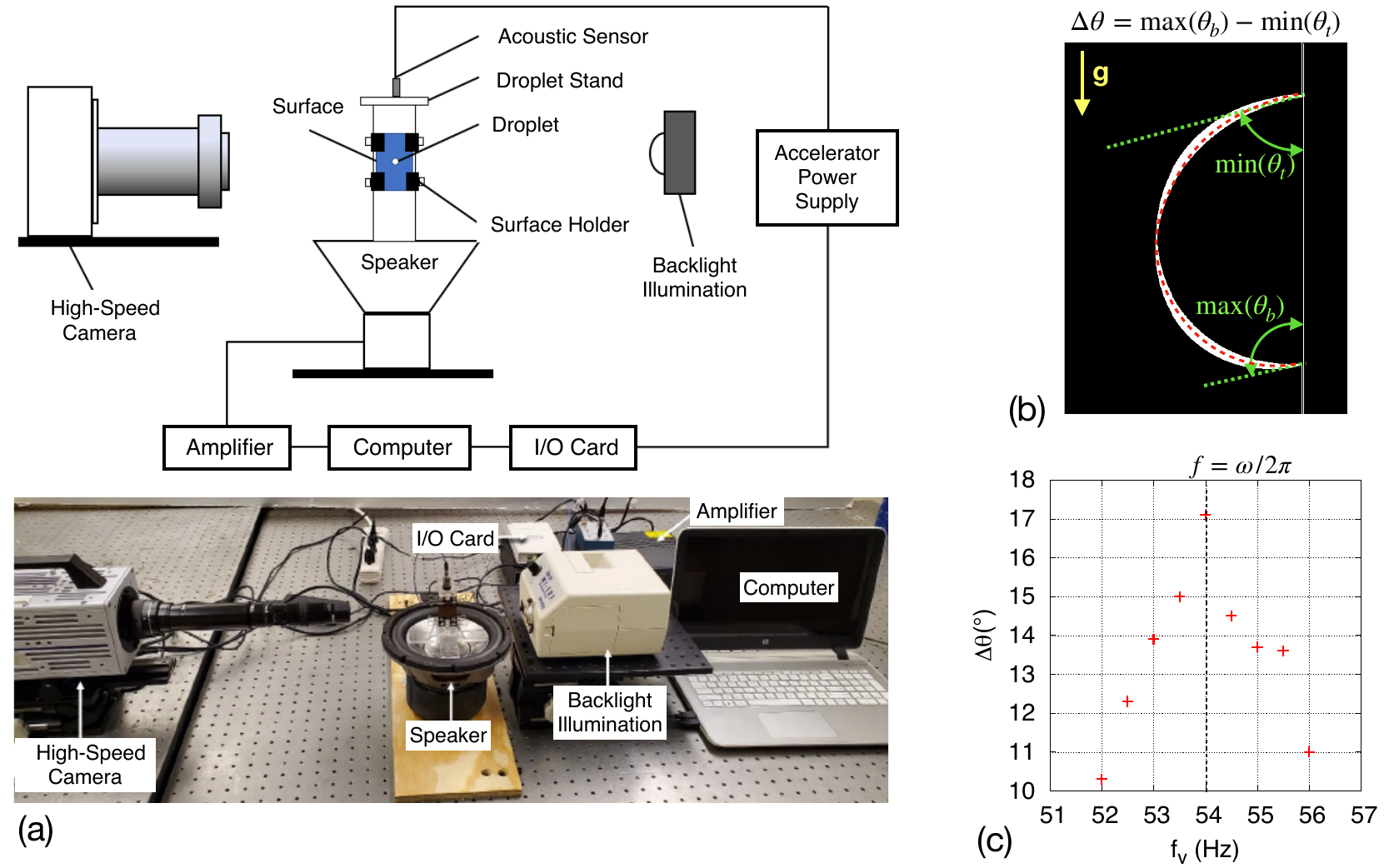}
 \caption{(a) Experiment setup. (b) Superimposition of drop edges on the central plane from 100 snapshots {for the polystyrene surface and drop volume 4.78 \textmu L. The red dashed line indicates the equilibrium shape. The contact angles at the top and bottom of the drop, $\theta_{t}$ and $\theta_{b}$, are measured. The maximum contact angle hysteresis is defined as $\Delta \theta =\max(\theta_b)-\min(\theta_t)$.} (c) Variation of $\Delta \theta$ as a function of the {surface} vibration frequency $f_v$. {The resonance angular frequency $\omega=2\pi f$ is identified as the peak of the function.}}
 \label{fig:experiment_setup}
\end{figure}

Experiments have also been performed to measure the oscillation frequency of the lateral mode. The drop oscillations were excited by the lateral vibration of the supported substrate. Distilled water drops were placed on the vertical substrate, which was fixed to a drop stand,  see figure~\ref{fig:experiment_setup}(a). 
The stand was attached to a sound speaker (Infinity Reference 860 W, Infinity Inc.) with an amplifier system (Russound P75-2 Channel Dual Source 75 W). The speaker provided sine wave vibrations of different frequencies to the substrate using a  connected online tone generator.  An accelerometer (352C04, PCB Piezoelectronics) was used to measure vibration frequencies and acceleration of the substrate. A National Instruments Data Acquisition System ((NI USB-5132) was used to interface the accelerometer with the computer. A high-speed camera (Photron SA3) with a halogen lamp (Fiber-Lite MI series, Dolan-Jenner Inc.) was used to capture the drop deformation. 

The different substrates and drop volumes tested are summarized in Table \ref{tab:experiment_cases}. Five different commercially available substrates with increasing $\theta_0$ were used. For each substrate, four different droplet volumes were tested. The wax substrate was prepared in-house by firmly attaching wax paper on polystyrene surface. All substrates, except for the wax one were cleaned with isopropyl alcohol, distilled water, and air dried before testing. The ranges of parameters considered in the simulation are wide enough to cover all the experimental cases. 

\begin{table}
  \begin{center}
  \begin{tabular}{ccc} 
        Substrate & $\theta_0$ (\textdegree) & $V_d$ (\textmu L)\\
        Al 6061 & 74.2 $\pm$ 1.80 & 3.30, 3.79, 4.29, 4.78 \\
        Cu 110 & 85.5 $\pm$  1.30 & 3.30, 3.79, 4.29, 4.78 \\
	Polystyrene & 93.0 $\pm$ 1.06   & 3.30, 3.79, 4.29, 4.78 \\
	PTFE        & 100.6 $\pm$ 0.70    & 3.30, 3.79, 4.29, 4.78 \\
	Wax   & 105.0 $\pm$ 0.60 & 3.30, 3.79, 4.29, 4.78 \\
    \end{tabular}
    \caption{Experimental cases. }
\label{tab:experiment_cases}
    \end{center}
\end{table}

For a given acceleration of the substrate, the excited drop oscillation amplitude is maximized when the substrate vibration frequency $\omega_v$ matches with the natural frequency $\omega$. The acceleration amplitude is controlled to be small to avoid drop sliding on the surface. Different $\omega_v$ were imposed,  and the range of $\omega_v$ tested was estimated based on the theoretical model of  \cite{Celestini_2006a} (C-K). For each $\omega_v$, 100 drop images were recorded. The drop contours on the central plane were identified using the Canny method in MATLAB and  superimposed as shown in figure~\ref{fig:experiment_setup}(b). {The contact angles at the top and bottom of the drops $\theta_t$ and $\theta_b$ are measured. The maximum contact angle difference, $\Delta \theta = \max(\theta_b)-\min(\theta_t)$, is used to characterize the excited drop oscillation amplitude.} The resonance frequency $\omega=2\pi f$ is then identified by the local maximum of $\Delta \theta$ as a function of $\omega_v=2\pi f_v$, see figure~\ref{fig:experiment_setup}(c). A small $f_v$ increment of 0.5 Hz were used near the local maximum for accurate measurement.

\section{Results and discussion}
\subsection{Oscillation frequency}
\begin{figure}
 \centering
 \includegraphics[width=0.65\textwidth]{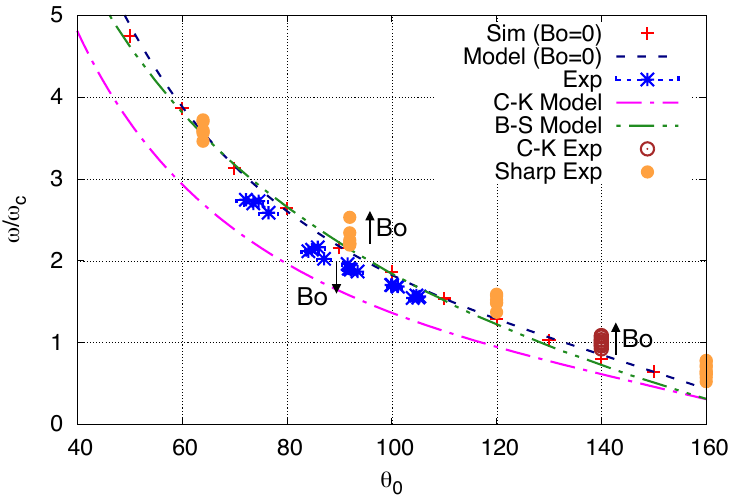}
 \caption{Present simulation, model, and experiment results for the frequency of the lateral oscillation mode as a function of the equilibrium contact angle $\theta_0$ for $\text{Bo}=0$, compared with previous models and experiments by \cite{Celestini_2006a} (C-K), \cite{Sharp_2011a}, and \cite{Bostwick_2014a} (B-S). {The experiments of C-K and Sharp are for sessile drops on a horizontal surface, for which the normalized frequency increases with Bo, while the present experiments are for a vertical surface, where the frequency decreases with Bo. The arrows indicates the variation of the normalized frequency as Bo increases. When Bo decreases, the experimental results for different surface orientations approach to the simulation and model results for $\text{Bo}=0$. }}
 \label{fig:Freq_Bo0}
\end{figure}

The simulation and experimental results for the oscillation frequency of the lateral mode $\omega$ are shown in figure~\ref{fig:Freq_Bo0}. The results of previous experiments by C-K and \cite{Sharp_2011a}, and theoretical models by C-K  and B-S are also included for comparison. It is observed that the experimental data for different drop volumes collapse approximately, when $\omega$ is normalized by $\omega_c$. This scaling relation $\omega\sim \omega_c$ indicates $\omega \sim V_d^{3/2}$, which is consistent with previous observations for minor gravity effect \citep{Noblin_2004a}. The normalized frequency $\omega/\omega_c$ for different $V_d$ still varies slightly due to the effect of Bo. For the experiments of C-K ($0.04\lesssim \text{Bo} \lesssim 0.19$) and Sharp ($0.16\lesssim \text{Bo} \lesssim 0.66$), the surface is horizontal, $\omega/\omega_c$ increases with Bo \citep{Sakakeeny_2021a}, while for the present experiment ($0.12\lesssim \text{Bo} \lesssim 0.16$) that uses vertical surfaces, $\omega/\omega_c$ decreases with Bo (as shown later in figure~\ref{fig:Freq_Bo}). {When Bo decreases, the experimental results for different surface orientations approach to the simulation results for $\text{Bo}=0$. }

The B-S model assumes Bo=0, so the equilibrium drop is a spherical cap. The inviscid linear stability theory is used to predict the frequency of the oscillation. The B-S predictions generally agree well with simulation results, see figure~\ref{fig:Freq_Bo0}. Small discrepancy are observed for $\theta_0\gg \ang{90}$, which may be due to the simplified contact-line boundary conditions used in the B-S model \citep{Sakakeeny_2020a}. The C-K model predicts the frequency by a different approach. The shape oscillation of the drop is modeled as a harmonic oscillator, so the frequency is $\omega=\sqrt{k/m_d}$, where $m_d$ is the drop mass. The effective spring constant $k$ is related to surface tension, which serves as a restoring force to bring the drop back to its equilibrium state. The C-K model is more convenient to use since an explicit expression of $\omega$ is given, but it significantly under-predicts the frequencies, as shown  in figure~\ref{fig:Freq_Bo0} and also in the original paper of C-K. 
%It was argued that the discrepancy may be due to the Stokes layer near the solid surface \cite{Celestini_2006a}. Nevertheless, as will be shown later, the discrepancy between the C-K model is mainly due to the missing effect of kinetic energy of fluid motion inside the drop. 

\subsection{An inviscid model to predict oscillation frequency}
\begin{figure}%
    \centering
       \includegraphics[width=0.95\textwidth]{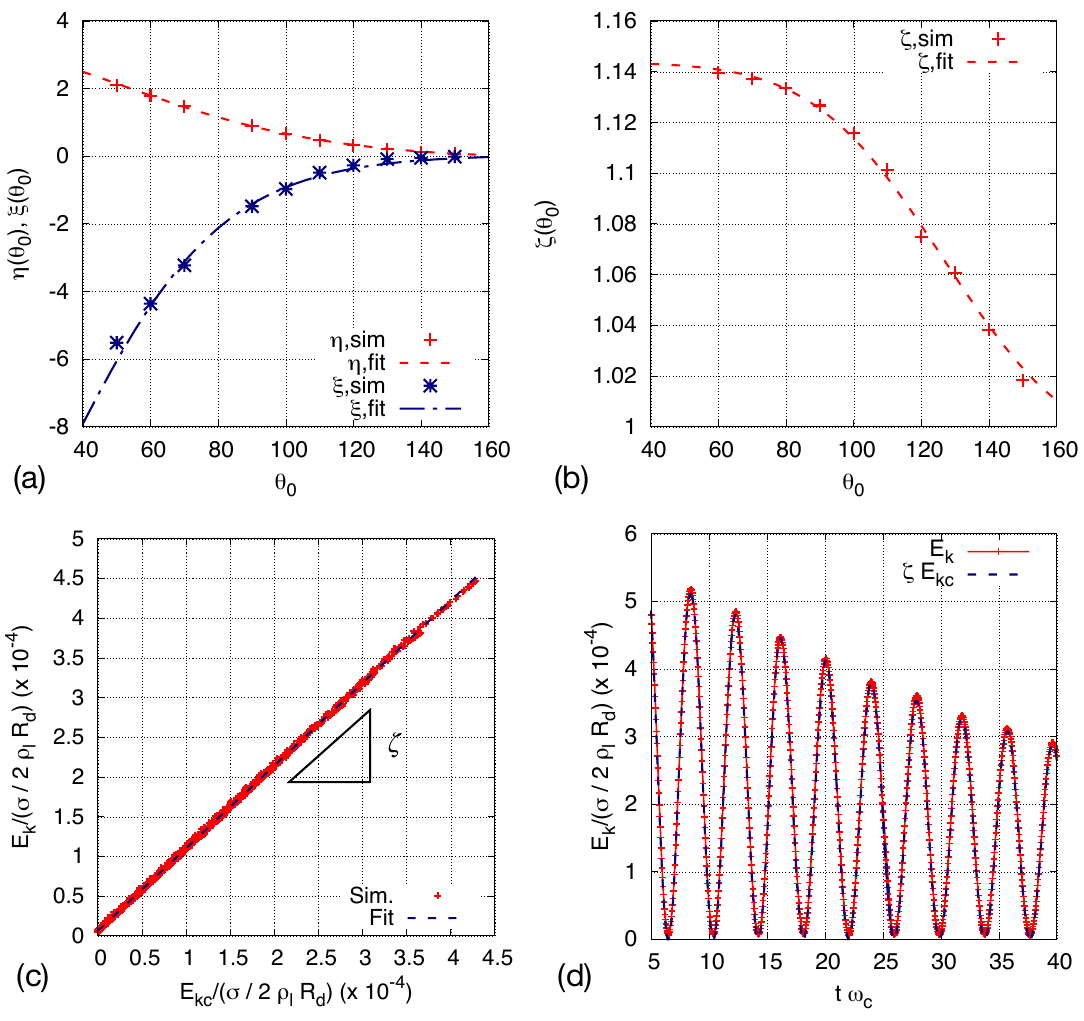}
    \caption{Model parameters (a) $\eta$ and $\xi$ and (b) $\zeta$ as functions of $\theta_0$. (c) $E_k$ as a function of $E_{kc}$ and (d) temporal evolution of $E_k$ and the approximation $\zeta E_{kc}$ for $\theta_0=\ang{140}$ and $\text{Bo}=0$.}
    \label{fig:Model_Para}%
\end{figure}

A new inviscid model is developed in this study. The goal is to achieve an explicit expression to accurately predict the oscillation frequency. Furthermore, the new model incorporates the effect of finite Bo, which is ignored in B-S and C-K models. Taking the equilibrium state for Bo=0 as the reference state, where the y-coordinate of the centroid and the drop surface area are $y_c= y_{c0}=0$ and $S=S_0$, respectively. The surface energy of the drop is  $E_s=\sigma(S-S_0)$. 
The surface area $S-S_0$ can be expanded as a function of $y_c-y_{c0}$, 
\begin{align}
	\frac{S-S_0}{S_0} = \eta\left(\frac{y_c-y_{c0}}{R_0}\right)^2 + \xi\left(\frac{y_c-y_{c0}}{R_0}\right)^4.
	\label{eq:surf_area}
\end{align}
Due to symmetry, only the even order terms remain and the terms higher than fourth order are truncated, assuming small oscillation amplitudes. The relation between $S$ and $y_c$ during oscillation is assumed to be similar to that for a static equilibrium drop under different body forces \citep{Celestini_2006a}. As a result, the parameters $\eta$ and $\xi$ depend only on $\theta_0$ and can be obtained from the solution for static drops. The \emph{Surface Evolver} code was used to obtain the static solutions for different body forces for a given $\theta_0$, which are then used to measure $\eta$ and $\xi$ following Eq.~\eqr{surf_area}. The procedures are similar to our previous study \citep{Sakakeeny_2020a} and thus are not repeated here. The results for $\eta$ and $\xi$ are shown in figure~\ref{fig:Model_Para}(a). For convenient use of the results, fitted correlations are provided, 
\begin{align}
	\eta(\theta_0)& = a_0 \left[ \exp \left(a_1 (\cos\theta_0+1) \right)-1\right]\, ,	
	\label{eq:eta}\\
	-\xi(\theta_0) &= b_0 \left[ \exp \left(b_1 (\cos\theta_0+1) \right)-1\right]\,, 
	\label{eq:xi}
\end{align}
where the fitting coefficients $[a_0,a_1]=[0.52.0.99]$ and  $[b_0,b_1]=[0.20,2.11]$. In the asymptotic limit $\theta_0\to \ang{180}$, both $\eta$ and $\xi$ approach zero, since in that limit the lateral mode does not induce deformation. The static solutions also indicate that $\eta>0$ and $\xi<0$ for all $\theta_0$. 

The gravitational potential energy is expressed as
$E_g=m_d g(y_c-y_{c0})$, and the overall potential energy is $E_p=E_s+E_g$. The equilibrium state is located at the minimum of $E_p$, namely $dE_p/dy_c=0$, which  yields
\begin{align}
	2\eta(y_{c1}/R_0) + 4\xi(y_{c1}/R_0)^3 + \text{Bo}_1=0\,,
	\label{eq:equil}
\end{align}
where $\text{Bo}_1 = mgR_0/\sigma S_0=0.42(2 + \cos \theta_0)^{1/3} (1-\cos \theta_0)^{-1/3} \text{Bo}$. The cubic equation can be solved for the equilibrium centroid position $y_{c1}$, near which $E_p$ can be expanded as 
\begin{align}
	E_p=\frac{1}{2}\ods{E_p}{y_c}(y_c-y_{c1})^2=\sigma S_0 \eta_1\left(\frac{y_c-y_{c1}}{R_0}\right)^2\,,
	\label{eq:potential_energy_expan}
\end{align}
where $\eta_1=\eta + 6 \xi (y_{c1}/R_0)^2$. 
%Assuming the small-amplitude oscillation around $y_{c1}$, the terms higher than second order are dropped. 
Though the exact solution exists for Eq.~\eqr{equil}, we use the approximation $y_{c1} \approx -{\text{Bo}_1}/{2\eta}$ by neglecting the cubic term. Then we get the approximation 
\begin{equation}
	\eta_1\approx\eta + \frac{3\xi}{2\eta^2} Bo_1^2\,.
\end{equation}

The kinetic energy of the drop fluid is $E_k=\rho_l \int_{V_d} |\bs{u}|^2/2\ dV$, {which is evaluated in the simulation  by the integral over the domain
$E_k = \rho_l \int f (u^2 + v ^2 + w^2) d V$, 
where $f$ is the liquid volume fraction, and $u$, $v$, and $w$ are the velocity components. } 
If the drop moves as a rigid body, the kinetic energy would be $E_{kc}=m_d v_c^2/2$, where $v_c=dy_c/dt$ is the y-component of the centroid velocity. Due to the internal flow with respect to the centroid induced by the shape oscillation, $E_k>E_{kc}$ \citep{Sakakeeny_2021a}. Furthermore, as shown in figure~\ref{fig:Model_Para}(c) that, $E_k$ is approximately a linear function of $E_{kc}$ when the drop oscillates. This is due to the fact that the temporal evolutions of energy of the internal flow and the magnitude of $v_c$ are in phase. As a result, the kinetic energy correction factor, defined as $\zeta=E_k/E_{kc}$, is approximately a constant in time and  can be measured from the simulation results as shown in figure~\ref{fig:Model_Para}(c). Then we can approximate
\begin{align}
	E_k\approx \zeta E_{kc} = \zeta m_d v_c^2/2\,.
	\label{eq:KE_approx}
\end{align}
The temporal evolutions of  $E_k$ and $\zeta E_{kc}$ are plotted in figure~\ref{fig:Model_Para}(b) and the latter is affirmed to be generally a good approximation for all time. Further tests show that $\zeta$ varies little over Bo and is only a function of $\theta_0$, see figure~\ref{fig:Model_Para}(b). The fitted function for $\zeta$ is given as 
\begin{align}
%	\zeta(\theta_0)&=\exp \left(c_1 (\cos\theta_0+1) \right)\, ,\\
	\zeta(\theta_0)& = c_1\mathrm{erf}\left[(\cos\theta_0+1)/c_2\right]+1 
	\label{eq:zeta}
\end{align}
where the fitting coefficients $[c_1,c_2]=[0.14,0.93]$. In the asymptotic limit $\theta_0\to \ang{180}$, $\zeta \to 1$ since there is no shape oscillation and the lateral mode corresponds to rigid-body translation of the drop.

The total energy $E=E_p+E_k$ is constant in time, so $dE/dt=0$. With Eqs.~\eqr{potential_energy_expan} and \eqr{KE_approx} we obtain
\begin{align}
	k(y_c-y_{c1})+m_d \ods{(y_c-y_{c1})}{t} = 0 \,,
	\label{eq:harmonic}
\end{align}
where $k=2\sigma S_0 \eta_1/(\zeta R_0^2)$. The oscillation frequency $\omega=\sqrt{k/m_d}$ can be expressed as
\begin{align}
	\frac{\omega^2}{\omega_c^2} = \frac{12 \left( \eta + \frac{3\xi}{2\eta^2} \text{Bo}_1^2\right)}{\zeta(2+\cos\theta_0)(1-\cos\theta_0)} \,.
	\label{eq:model}
\end{align}

%\subsection{Model predictions for $\text{Bo}=0$}
In the limit of $\text{Bo}=0$, Eq.~\eqr{model} reduces to 
\begin{align}
	\frac{\omega^2}{\omega_c^2} = \frac{12 \eta}{\zeta(2+\cos\theta_0)(1-\cos\theta_0)} \,.
	\label{eq:model_Bo0}
\end{align}
The results of Eq.\ \eqr{model_Bo0} are plotted in figure~\ref{fig:Freq_Bo0}, which are shown to agree remarkably well with the simulation results. The C-K model for Bo=0 can be written as
\begin{align}
	\frac{\omega_{CK}^2}{\omega_c^2} = \frac{6 \eta}{(2+\cos\theta_0)(1-\cos\theta_0)} \,.
	\label{eq:CK-model}
\end{align}
Comparing the Eqs.~\eqr{model_Bo0} and \eqr{CK-model} for $\text{Bo=0}$, it is shown that the present model yields better predictions due to the incorporation of contribution of the kinetic energy of the internal flow through $\zeta$. 

\subsection{Model predictions for finite Bond numbers}
Since numerical solutions are required for the model parameters $\eta$, $\xi$ and $\zeta$, the present model, similar to the C-K model, is semi-analytical. Nevertheless, once these parameters are established, as shown in Eqs.~\eqr{eta},\eqr{xi},\eqr{zeta}, the explicit expression Eq.~\eqr{model} is ready to predict the natural frequencies for the lateral oscillation of drops pinned on a vertical surface with different $\theta_0$ and Bo. 

\begin{figure}%
    \centering
       \includegraphics[width=0.95\textwidth]{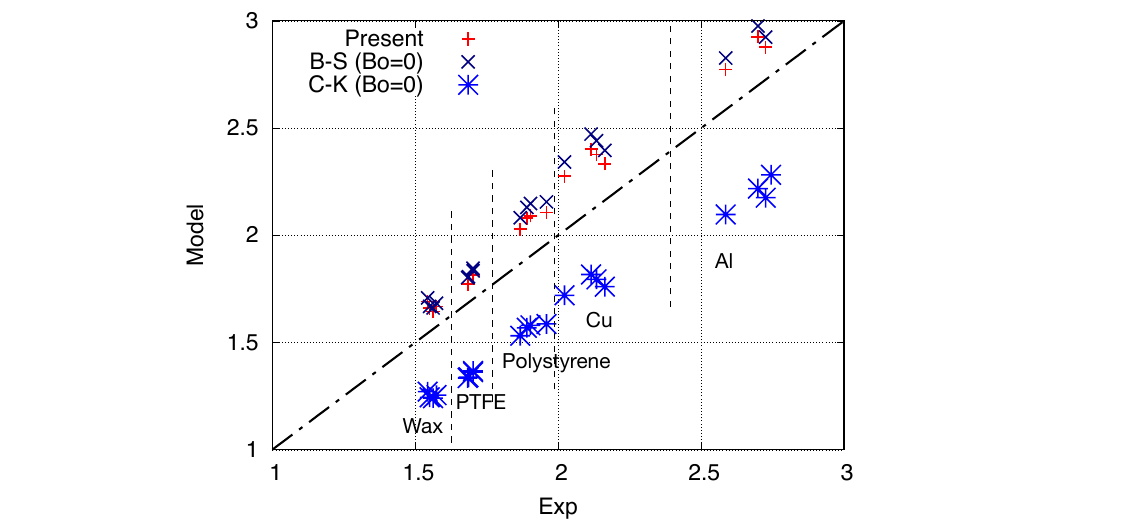}
    \caption{{Comparison between experimental results of the normalized frequency $\omega/\omega_c$ and model predictions with the present, B-S and C-K models. }}
    \label{fig:Model_exp}%
\end{figure}

{
The predictions of $\omega/\omega_c$ by the present, B-S, and CK models are compared with the present experimental results for finite Bo in Table \ref{tab:experiment_model}. The results are also plotted in figure \ref{fig:Model_exp}. Since the C-K model neglects the effect of kinetic energy of the fluid motion within the drop, it significantly under-predicts the frequency, up to 21\%. In contrast, the B-S model, which neglects the effect of Bo, over-predicts the frequency, up to 17\%. As shown in figure \ref{fig:Freq_Bo0}, the present model yields very similar results as the B-S model for $\text{Bo}=0$ in the range of $\theta_0$ considered in the present experiments. By incorporating the contribution of Bo, the present model yields more accurate predictions than the other two models for all cases. The relative errors are less than 12\% for all cases. Excluding the cases for copper, the errors are less than 9\%. 
}

{
The present model over-predicts the frequency for all cases, compared to the experimental results. There are a couple potential reasons that contribute to the discrepancy between the model and experimental results. Similar to the B-S and C-K models, the contact area is taken to be a  circle with radius as $R_c=R_0\sin{\theta_0}$ in the present model. This ideal contact area corresponds to the spherical-cap shape of the drop when gravity is absent. The contact area is then fixed during oscillation, since the contact line is pinned. In the experiment, the contact area may be slightly smaller than ideal contact area considered in the model, since gravity is present when the drop is deposited on the surface. The smaller contact area will result in weaker constraint on the drop and a lower oscillation frequency. Furthermore, though in the experiment the oscillation amplitude is controlled to be very small, the contact line may still not be fully pinned and may have oscillated in a very small amplitude near the equilibrium position. In contrast, the contact line is perfectly pinned in the model and simulation. The mobility of the contact line will result in a reduction of the oscillation frequency for the same $\theta_0$ and $\text{Bo}$, according to our previous study \citep{Sakakeeny_2021a}. 
}

\begin{table}
  \begin{center}
  \begin{tabular}{ccccccccccc} 
        Substrate & $V_d$ (\textmu L) & $\theta_0$ (\textdegree) & Bo & $(\frac{\omega}{\omega_c})_{\text{exp}}$ & $(\frac{\omega}{\omega_c})_{\text{mod}}$ & $\varepsilon_{\text{mod}}$ (\%)& $(\frac{\omega}{\omega_c})_{\text{BS}}$ & $\varepsilon_{\text{BS}}$ (\%) & $(\frac{\omega}{\omega_c})_{\text{CK}}$ & $\varepsilon_{\text{CK}}$ (\%) \\
        Al 6061 	& 3.30 & 74.60 	& 0.116 & 2.726	& 2.878	& 5.3 & 2.924	& 7.3		& 2.175 	& -20.2\\
        			& 3.79 & 74.50 	& 0.127 & 2.586	& 2.773	& 6.7 & 2.827	& 9.3		& 2.097 	& -18.9\\
        			& 4.29 & 72.15 	& 0.138 & 2.745	& 3.013	& 8.9 & 3.054	& 11.3	& 2.282 	& -16.9\\
        			& 4.78 & 73.60	& 0.149 & 2.699	& 2.926	& 7.8 & 2.976	& 10.2	& 2.218 	& -17.8\\
        Cu 110 	& 3.30 & 85.85 & 0.116 & 2.163		& 2.333	& 7.3 & 2.396	& 10.8	& 1.760 	& -18.7\\
        			& 3.79 & 84.80 & 0.127 & 2.134	& 2.376	& 10.2 & 2.441	& 14.4	&1.794 	& -15.9\\
        			& 4.29 & 87.10 & 0.138 & 2.022	& 2.276	& 11.2 & 2.342	& 15.8	& 1.720 	& -14.9\\
        			& 4.78 & 84.10	& 0.149 & 2.115	& 2.401	& 11.9 & 2.472	& 16.9	& 1.817 	& -14.1\\
     Polystyrene 	& 3.30 & 91.55 & 0.116 & 1.959		& 2.106	& 7.0 & 2.156	& 10.0	& 1.586 	& -19.0\\
        			& 3.79 & 92.20 & 0.127 & 1.890	& 2.080	& 9.1 & 2.130 	& 12.7	& 1.568 	& -17.0\\
        			& 4.29 & 93.50 & 0.138 & 1.867	& 2.029	& 8.0 & 2.082	& 11.5	& 1.531 	& -18.0\\
        			& 4.78 & 91.75	& 0.149 & 1.903	& 2.091	& 9.0 & 2.148	& 12.9	& 1.580 	& -16.9\\
	PTFE 	& 3.30 & 100.10 & 0.116 & 1.702	& 1.812	& 6.1 & 1.834	& 7.8		& 1.361 	& -20.2\\
        			& 3.79 & 99.85 & 0.127 & 1.702	& 1.817	& 6.4 & 1.846	& 8.5		& 1.367 	& -19.7\\
        			& 4.29 & 101.20 & 0.138 & 1.683	& 1.772	& 5.0 & 1.802	& 7.0		& 1.334 	& -20.7\\
        			& 4.78 & 101.05 & 0.149 & 1.686	& 1.773	& 4.9 & 1.807	& 7.2		& 1.338 	& -20.6\\
        Wax 		& 3.30 & 105.10 & 0.116 & 1.550	& 1.660	& 6.8 & 1.668	& 7.6		& 1.244 	& -19.7\\
        			& 3.79 & 103.95 & 0.127 & 1.543	& 1.691	& 8.8 & 1.708	& 10.7 	& 1.270 	& -17.7\\
        			& 4.29 & 104.65 & 0.138 & 1.574	& 1.667	& 5.6 & 1.683	& 6.9		& 1.254 	& -20.3\\
        			& 4.78 & 105.30 & 0.149 & 1.562	& 1.644	& 5.0 & 1.661	& 6.3		& 1.240 	& -20.6\\
    \end{tabular}
    \caption{{Comparison between experimental results of the normalized frequency $\omega/\omega_c$ and model predictions with the present, \cite{Bostwick_2014a} (B-S) and \cite{Celestini_2006a} (C-K) models.} }
\label{tab:experiment_model}
    \end{center}
\end{table}

{In the experiment, Bo is varied only in a very small range ($0.12< \text{Bo} < 0.15$). Drops with larger Bo are very unstable and may slide even when the surface vibration amplitude is very small. To better pin the contact line and to enable experiment for larger Bo, a more sophisticated surface treatment, such as that by \cite{Chang_2015a}, is required, which will be relegated to future work. In order to validate the model for higher Bo, the model predictions are compared with the fully-resolved simulation results}. 

\begin{figure}%
    \centering
       \includegraphics[width=0.95\textwidth]{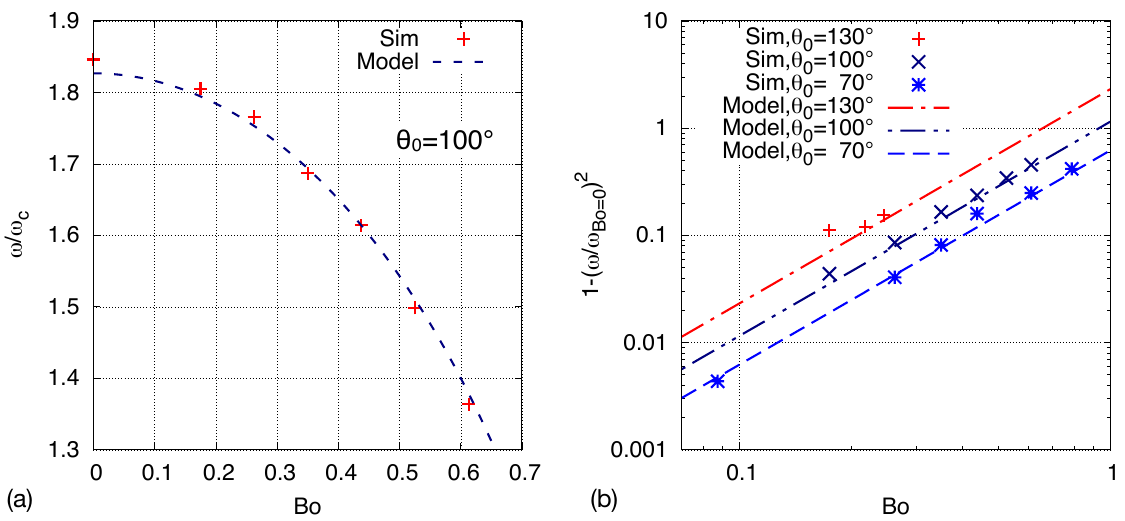}
    \caption{{Comparison between the model and simulation results for }oscillation frequency as a function of Bo and $\theta_0$.  }
    \label{fig:Freq_Bo}%
\end{figure}

The model and simulation results of the normalized frequency $\omega/\omega_c$ as a function of Bo for $\theta_0=\ang{100}$ are shown in figure~\ref{fig:Freq_Bo}(a). It is seen that $\omega/\omega_c$ decreases with Bo and excellent agreement between the present model and the simulation results is observed. {It is conventionally considered that the effect of gravity on the the supported drop oscillation is negligible if $\text{Bo}<1$, namely when the drop radius is smaller than the capillary length \citep{Chang_2015a}. However, it can be seen that the frequency decreases for more than 25\% when $\text{Bo}$ increases from 0 to 0.62. }

{The variation of $\omega/\omega_c$ over Bo can be better depicted if $\omega$ is normalized by its $\text{Bo}=0$ counterpart $\omega_{\text{Bo}=0}$, From Eq.~\eqr{model}, it can be shown that}
\begin{align}
	{\omega^2}/{\omega_{\text{Bo}=0}^2} = 1+ ({3\xi}/{2\eta^3}) \text{Bo}_1^2\,.
	\label{eq:model_Bo}
\end{align}
{The equilibrium drop results show that $\xi<0$ for all $\theta_0$}  (see figure \ref{fig:Model_Para}(a)), indicating that that the curvature of the $S$-$y_c$ curve (Eq.~\eqr{surf_area}) decreases as $y_c$ increases. As a result, $\omega/\omega_c$ decreases with Bo, see figure~\ref{fig:Freq_Bo}(a). Furthermore, the leading correction term for ${\omega^2/\omega_{\text{Bo}=0}^2}$ scales with $\text{Bo}^2$. Therefore, for finite and small Bo, $\omega/\omega_c$ decreases quadratically with Bo. It is difficult to vary Bo for a wide range in experiment to examine the scaling relation, but that can be done in the simulation. We have change $g$ from 0 to 0.9 m/s$^2$, yielding $0\le\text{Bo}\le 0.78$. The model  and simulation results for $\theta_0=\ang{70}, \ang{100}$, and $\ang{130}$ are plotted in figure~\ref{fig:Freq_Bo}(b). Since drops with large $\theta_0$ become unstable for large Bo, the range of Bo for $\theta_0=\ang{130}$ is smaller than other cases. The model predictions (Eq.~\eqr{model_Bo}) agree very well with the simulation results, validating the quadratic scaling law revealed in the model. 

In the present study, Bo is varied by changing the gravity $g$. Therefore, the results imply that the natural  frequency of the lateral oscillation mode will increase if the drops oscillations occur in a reduced gravity environment. This knowledge is important to the surface-vibration dropwise condensation applications in the space station. 

The present model can also be used to estimate the frequencies for the drops of different volumes. For a given gravity, Bo increases with $V_d$. As a result, $\omega/\omega_c$ decreases as $V_d$ increases. Since the model is based on an inviscid assumption and the viscous effect on the oscillation frequency \citep{Lamb_1932a, Prosperetti_1980a} is not accounted, it is valid only for $\text{Oh}\ll 1$. For low-viscosity liquids like water, Oh for millimeter-drops is generally small. Yet, as Oh increases as the drop size decreases,  caution is required if the model is applied to very small drops. 

\section*{Acknowledgement}
This work was supported by the startup fund at Baylor University and the National Science Foundation (1942324). The Baylor High Performance and Research Computing Services (HPRCS) have provided the computational resources that have contributed to the research results reported. The authors also thank Prof.~St\'ephane Zaleski for the helpful discussions on numerical model of pinned contact lines. 

%\section*{Data Availability}
%\vspace{0.1in}
%The data that support the findings of this study are available from the corresponding author upon reasonable request.

\section*{Declaration of Interests}
The authors report no conflict of interest.

%\bibliographystyle{jfm}
%\bibliography{refs}

\end{document}